# Predicting the Early Stages of Solid-State Precipitation in Al-rich Al-Pt Alloys

Shenghan Su [a], Laure Bourgeois [a, b, *], Nikhil V. Medhekar [a, *]

[a] *Department of Materials Science and Engineering, Monash University, VIC, 3800, Australia.*

[b] *Monash Centre for Electron Microscopy, Monash University, VIC, 3800, Australia.*

[*] *Corresponding authors.*

*E-mail: nikhil.medhekar@monash.edu; laure.bourgeois@monash.edu*

**Abstract**

The high strength of structural aluminium alloys depends strongly on the controlled precipitation of specific intermetallic phases, whose identity and crystal structure can be difficult to predict. Here, we investigate the Al-Pt system, which shows some similarity to the Al-Cu system as one of their main intermetallic phases, $Al_2Pt$, is nearly isostructural with $\theta'$ ($Al_2Cu$), the metastable phase responsible for the high-strength of Al-Cu alloys. However, phases in Al-Pt alloys are complex and have not been studied in detail. Using a combination of density-functional theory (DFT) calculations and classical nucleation theory (CNT) applied to the Al-Pt system, we design a workflow to predict the thermodynamics of solid solution, intermediate phases such as GP zones, stable and metastable precipitates, and their precipitation sequence. This workflow can be applied to an arbitrary binary alloying system. We confirm the known stable phases $Al_4Pt$, $Al_{21}Pt_8$, $Al_2Pt$, $Al_3Pt_2$, $AlPt$ ($\alpha$ & $\beta$), $Al_3Pt_5$, $AlPt_2$ ($\alpha$ & $\beta$) and $AlPt_3$ ($\alpha$ & $\beta$). We also reveal the possible existence of two phases of chemical formulae $Al_5Pt$ and $Al_3Pt$. This large number of intermetallic phases is due to the strong bonding between Al and Pt, which also leads to significant favourable Pt solute formation energy in the Al matrix. Our findings are compared with the known precipitation characteristics of the binary Al-Cu and Al-Au systems. We find that the $\theta'$-like $Al_2Pt$ precipitate phase has a lower coherent interfacial energy than $\theta'$. Our calculations strongly suggest that $Al_2Pt$ will precipitate first in Al-rich Al-Pt alloys and will form bulk-like interfaces similar to $\eta$ ($Al_2Au$) rather than like $\theta'$ ($Al_2Cu$).

# 1 Introduction

Controlled solid-state precipitation is commonly used for achieving significant strengthening in light structural alloys. This is because precipitates of appropriate shape, distribution, number density and crystallography can constitute effective barriers to the movement of dislocations. In general, high-strength alloys require uniformly and densely distributed precipitates with high aspect ratios and a rational crystallographic orientation [1]. A well-known alloy system containing such precipitates is the Al-Cu system, which forms the basis of an important class of engineering alloys for aircraft and aerospace applications [2]. In Al-Cu alloys, a fine and uniform precipitation of $\theta''$ ($Al_3Cu$) and $\theta'$ ($Al_2Cu$) precipitates with aspect ratios greater than 50:1 can lead to tensile strength of up to 500 MPa [2]. The $\theta'$ phase is metastable and its crystal structure bears little resemblance to known bulk phases in the Al-Cu system [3]. More generally, precipitates forming in the early stages of precipitation are commonly metastable phases not expected to form based on the bulk phase diagram (e.g., $\eta'$ (AlCu) in Al-Cu alloys [4]) or they exhibit distorted structures of known bulk phases (e.g. $\Omega$ phase in Al-Cu alloys [5]). Consequently, predicting the early stages of precipitation in a given alloy system through atomic-scale modelling rather than in ad-hoc and empirical way, is a very challenging task. Yet this endeavour is of great importance for discovering and enhancing the properties of structural Al alloys. This challenge has motivated us to investigate the solid-state precipitation in the Al-rich part of the Al-Pt alloy system using advanced atomistic simulation methods.

In the periodic table, Pt sits next to the column of Cu and Au, and shares their cubic close-packed structure. The Al-Pt system exhibits several features that should make it interesting to compare with better known Al-Cu and Al-Au binary alloys containing plate-shaped precipitates. Firstly, among all Al binary alloys, the $Al_2Pt$ phase is the only stable θ′-like intermetallic phase besides η ($Al_2Au$) [6]. Secondly, several gaps remain in our knowledge of the precipitation characteristics of Pt-containing aluminium alloys. The latest research on the Al-Pt system indicates the following phases to be stable: $Al_4Pt$, $Al_{21}Pt_8$, $Al_2Pt$, $Al_3Pt_2$, AlPt (α & β ), $Al_3Pt_5$, $AlPt_2$ (α & β ) and $AlPt_3$ (α & β ) [7-9]. Here, α is the low-temperature polymorph and β is the high-temperature one—see Table 1 for details. $Al_2Pt$ has a cubic $CaF_2$-type structure with $a_{Al_2Pt} = 5.91$ Å [10], which is similar to the crystal structure of θ′ ($Al_2Cu$) [3] when viewed along <110> as shown in Figure 1. There remain many unknowns regarding the Al-Pt alloy system: recent experimental and computational studies have focused on the determination of bulk intermetallic phases [7, 8, 11-13] and their applications [14, 15], but have not addressed the thermodynamics of solid solution, precipitate interfacial structures and precipitation sequences. One particularly intriguing feature of the Al-Pt binary system is the existence of several stable and metastable phases of lower Pt concentration than $Al_2Pt$ [9], which is contrary to the Al-Cu [16] and Al-Au [17] systems where no stable Al-rich phases exist before the θ′ and η phases. However, there is hitherto no consensus on the precipitation sequences in the Al-Pt system. For instance, it remains unclear as to which phase will form after quenching an Al-Pt alloy, and which one will precipitate first [18-20].

In this work, we used first-principles calculations based on density-functional theory (DFT) combined with classical nucleation theory (CNT) to determine the thermodynamical stabilities of different phases, precipitate interfacial energies, and precipitation sequences in the Al-Pt binary system. We reveal the possible existence of two new phases, $Al_5Pt$ and $Al_3Pt$[7-9]. Among all phases, only θ′-like $Al_2Pt$ is predicted to form thermodynamically stable coherent interfaces along the $\{001\}_{Al}$ planes and is likely to precipitate first directly from the supersaturated solid solution, thereby bypassing phases with higher Al content which could be expected to form first based on chemical potential alone. We illustrate that solute clusters, GP zones, and non-bulk like interfaces are unlikely to form due to the significant low solute formation energy of Pt in the Al matrix. Our study thus fills knowledge gaps in the Al-Pt alloy system and more generally, provides a workflow to determine precipitate phases and precipitation in an arbitrary binary alloying system.

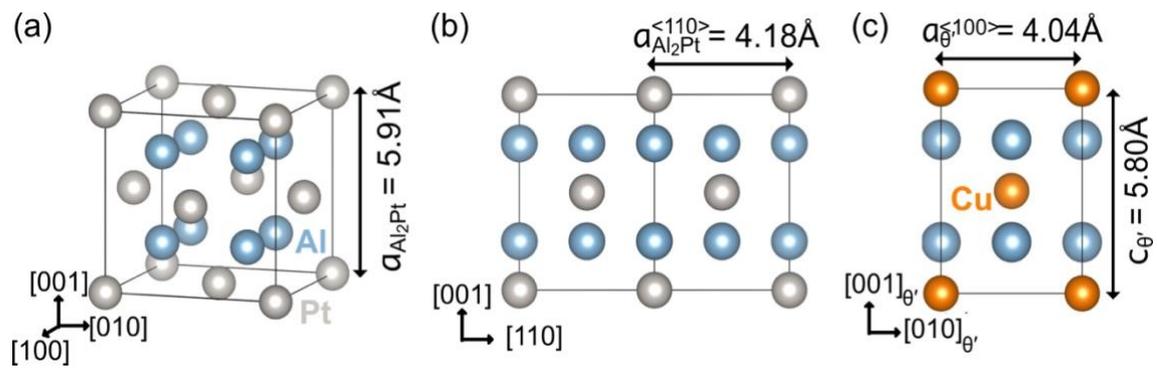

**Figure 1.** Experimentally determined crystal structure of (a) cubic $Al_2Pt$ and (b) when it is viewed along <110>. (c) Tetragonal $\theta'$ ($Al_2Cu$) viewed along <100>.

## 2 Methods

We first calculate the relaxed lattice parameters, thermodynamic characteristics, elastic properties, and interfacial energies of phases in the binary Al-Pt system using density-functional theory (DFT). These DFT results are then adopted as input parameters for the classical nucleation theory (CNT) calculations to examine the precipitation sequences.

All DFT calculations were carried out using Vienna ab initio simulation package (VASP) [21-24]. The plane-wave basis set was described by the projector augmented wave method (PAW) [25, 26]. The Perdew-Burke-Ernzerhof (PBE) exchange-correlation functional for the generalized-gradient approximation (GGA) were used [27]. For Al, Pt, Cu, Au and Ag, $3s^23p^1$, $5d^96s^1$, $3d^{10}4s^1$, $5d^{10}6s^1$, and $4d^{10}5s^1$ were treated as valence electrons respectively. The cut-off energy of 500 eV and the grid spacing of 0.08 /Å were used to generate a gamma-centred Monkhorst-Pack grid for sampling the Brillouin-zone to achieve sufficient accuracy for all calculations. To achieve a sufficient accuracy, energies and forces were converged within $1 \times 10^{-6}$ eV and f 0.01 eV/Å t, respectively. For calculations of phonons, we used a tighter energy tolerance criterion of $1 \times 10^{-8}$ eV.

We applied the CNT formalism to calculate the thermodynamics of precipitate nucleation:

$$\Delta G_{tot} = V\Delta G_{chem} + V\Delta G_{el} + A\gamma, \qquad (1)$$

where $V$ is the volume of the nucleus, $\Delta G_{chem}$ is the chemical energy, $\Delta G_{el}$ is the elastic strain energy, $\gamma$ is the average interfacial energy, and $A$ is the area of the interfaces between the precipitated phases and matrix. Further details of CNT calculations are presented in the Supplementary Information 1.

# 3 Results

We first present the thermodynamic characteristics of various phases in Al-Pt binary system as calculated by DFT methods (see Sec. 3.1, 3.2, and 3.3). Based on the obtained thermodynamic characteristics, we propose phases that can possibly precipitate from the supersaturated solid solution of Pt in Al matrix. Next we investigate interfacial structure, interfacial energies and elastic properties of possible precipitated phases in Sec. 3.4, 3.5, and 3.6 respectively. Finally, we estimate the critical nucleation radii and nucleation barriers based on computed energies via CNT (see Sec. 3.7).

## 3.1 Stable and Metastable Phases

We first investigate intermetallic phases between Al and Pt to provide a preliminary understanding of the binary Al-Pt system. Here we investigated the crystal structure and stability of the well accepted as well as controversial phases in this alloy system. Our results and data reported in literature are summarised in Table 1. A challenge in determining thermodynamic characteristics of Al-rich Al-Pt alloy system is that the full crystal structures of some reported phases, such as $Al_5Pt$ [18] and $Al_3Pt$ [7, 19, 20], are still unknown. We thus modelled these phases based on known intermetallic phases of the same chemical formula in binary systems with elements in the same group of Al and Pt. Among all possible crystal structures, those with the lowest formation energy are selected. $Al_5Pt$ is modelled from $Ga_5Pd$ and $Al_3Pt$ is modelled from $Tl_3Pt$. Further details for these calculations are provided in Supplementary Information 2. We assess the stability or metastability of different Al-Pt phases by the formation energy per atom ($E_f^{at.}$), calculated as

$$E_f^{at.} = \frac{E_{bulk}^{Al_xPt_y} - xE_{bulk}^{Al} - yE_{bulk}^{Pt}}{x+y}, \qquad (2)$$

where $E_{bulk}^{Al_xPt_y}$ is the total energy of a bulk phase containing $x$ Al and $y$ Pt atoms, $E_{bulk}^{Al}$ and $E_{bulk}^{Pt}$ are the per atom energies of bulk Al and Pt, respectively. A more negative formation energy represents a higher stability of the concerned phase. Since the precipitation in binary Al alloys occurs at temperatures in the range of 100-300 °C, we have also calculated the Helmholtz free energy of formation per atom ($F_f^{at.}$), similarly to LLorca et al.'s definition [28]

$$F_f^{at.} = \frac{F_{bulk}^{Al_xPt_y} - xF_{bulk}^{Al} - yF_{bulk}^{Pt}}{x+y}, \qquad (3)$$

The corresponding Helmholtz free energy ($F$) for each case is calculated as

$$F = E - TS, \qquad (4)$$

where $E$ is the energy calculated at 0 K and $S$ is the vibrational entropy contribution as a function of temperature. Again, a more negative free energy of formation represents greater stability. Here we have only considered vibrational contributions to the total entropy, since the configurational entropy can be neglected due to the low content of Pt, and the electronic entropy is small compared to vibrational entropy even for transition metals [29].

Figure 2 shows the calculated formation energies of intermetallic phases in the binary Al-Pt alloy system. These agree well with experimental measurements [9], except for an outlier at a concentration of 74.7 at. % Pt. The Al$_5$Pt phase possesses a formation energy close to the convex hull, which is in agreement with a previous DFT calculation [11]. Therefore, Al$_5$Pt is a possible stable phase. Figure 3 shows changes in the Helmholtz free energy of formation per atom from 0 $K$ to 1000 $K$ of selected Al-rich phases in the Al-Pt binary system. Whereas most phases become less stable at high temperatures, $Al_3Pt$ shows the opposite behaviour because it is stabilised by entropy at high temperatures. Therefore, $Al_3Pt$ is a possible stable phase at high temperatures. The $Al_5Pt$ phase has a small energy increment with temperature, supporting the previous argument about its stability.

**Table 1.** Crystal structures and formation energies of selected phases in the binary Al-Pt system as calculated by DFT methods in the present work. The values with superscripts are the experimental values reported in previous studies.

| Formula | Space group | Lattice parameters (Å) | | | $E_f^{at.}$ (eV/atom) |
|---|---|---|---|---|---|
| | | $a$ | $b$ | $c$ | |
| $Al_5Pt$ | I4/mcm | 6.40 | 6.40 | 10.03 | -0.43 |
| $Al_4Pt$ | P3c1 | 13.14 | 13.14 | 9.72 | -0.54 |
| | | 13.08 [a] | 13.08 [a] | 9.63 [a] | |
| $Al_3Pt$ | Pm$\bar{3}$n | 4.98 | 4.98 | 4.98 | -0.51 |
| $Al_{21}Pt_8$ | I4$_1$/a | 13.05 | 13.05 | 10.76 | -0.76 |
| | | 12.95 [b] | 12.95 [b] | 10.66 [b] | |
| $Al_2Pt$ | Fm$\bar{3}$m | 5.94 | 5.94 | 5.94 | -0.89 |
| | | 5.91 [b] | 5.91 [b] | 5.91 [b] | |
| $Al_3Pt_2$ | P$\bar{3}$m1 | 4.21 | 4.21 | 5.17 | -0.97 |
| | | 4.21 [b] | 4.21 [b] | 5.17 [b] | |
| α-AlPt | P2$_1$3 | 4.91 | 4.91 | 4.91 | -1.04 |
| | | 4.87 [b] | 4.87 [b] | 4.87 [b] | |
| β-AlPt | Pm$\bar{3}$m | 3.09 | 3.09 | 3.09 | -0.93 |
| | | 3.13 [b] | 3.13 [b] | 3.13 [b] | |
| $Al_3Pt_5$ | Pbam | 5.47 | 10.87 | 4.00 | -0.91 |
| | | 5.41 [b] | 10.73 [b] | 3.95 [b] | |
| α-$AlPt_2$ | Pmma | 5.50 | 3.95 | 16.48 | -0.86 |
| | | 5.43 [b] | 3.92 [b] | 16.31 [b] | |
| β-$AlPt_2$ | Pnma | 5.46 | 4.11 | 7.99 | -0.85 |
| | | 5.40 [b] | 4.05 [b] | 7.90 [b] | |
| α-$AlPt_3$ | P4/mbm | 5.51 | 5.51 | 7.94 | -0.70 |
| | | 5.45 [c] | 5.45 [c] | 7.82 [c] | |
| β-$AlPt_3$ | Pm$\bar{3}$m | 3.92 | 3.92 | 3.92 | -0.68 |
| | | 3.88 [b] | 3.88 [b] | 3.88 [b] | |

[a, b, c] Experimental values reported in [8], [10], and [30] respectively.

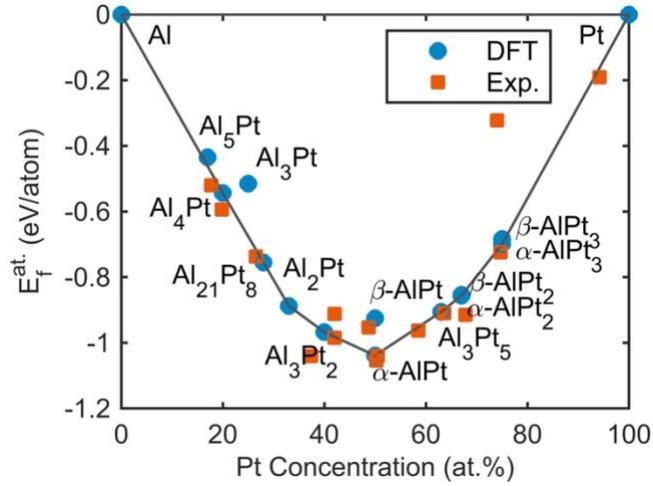

**Figure 2.** Formation energies of intermetallic phases in the binary Al-Pt system as calculated by DFT (blue circles) and compared to experimental measurements (red squares). There is a good agreement between experiments from previous studies and our calculations. The solid line represents a convex hull plotted based on DFT calculations.

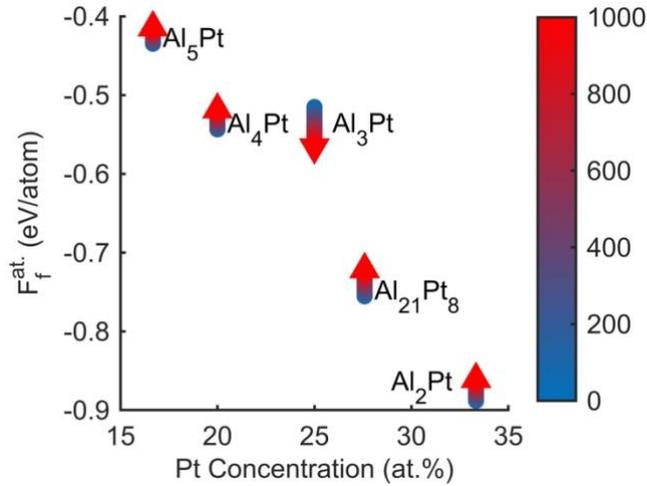

**Figure 3.** Change in the Helmholtz free energy of formation from 0 $K$ to 1000 $K$ of selected Al-rich phases in the Al-Pt binary system. Whereas most phases become less stable at high temperatures as expected, Al$_3$Pt shows the opposite behaviour.

## 3.2 The Pt Solid Solution

Since we are interested in predicting precipitation behaviour of Al-rich precipitate phases from a supersaturated solid solution of Pt in Al matrix, the natural starting point is the determination of Pt solute formation energies and its solubility. To interpret the results, we also examine the nature of the electronic interactions between solute and matrix. The solute formation energy per solute ($E_f^{s.}$) can be calculated by

$$E_f^{s.\ X} = E_{bulk}^{Al_{N-1}X} - \frac{N-1}{N} E_{bulk}^{Al_N} - E_{bulk}^{X}. \tag{5}$$

Here, a more negative value indicates a more stable state of solute X in Al matrix. We use a $4 \times 4 \times 4$ supercell to avoid any interactions between solute atoms, and N is the number of atoms in the supercell. Because Pt shows certain similarities to Cu and Au in the Al matrix as mentioned, we compare the solute formation energy of Pt to that of group 11 elements including Cu, Ag and Au (see Table 2). To our knowledge, no previous work exists for Pt. Our calculated results for other elements compare well with previous DFT calculations [31], and the minor discrepancies may be due to the smaller supercell size used in previous work.

    We find that Pt has the most negative formation energy among all selected solutes, indicating Pt solute should be stable in Al matrix. However, the solute formation energy is not directly correlated to the experimental solubility limit (see Table 2). This is because ordered phases will form beyond the solubility limit, and these phases are easier to form when two elements bind strongly according to the Hume-Rothery rules [32]. In addition, the formation energy is calculated at 0K and misses the contribution from the entropy at finite temperatures. Therefore, to evaluate the solubility of Pt in Al matrix, we calculate its solubility limit ($c_s$) as [33]

$$c_s = \exp\left(\frac{-\Delta F^{S.}_{excess}}{k_B T}\right), \qquad (6)$$

$$\Delta F^{S.}_{excess} = \frac{F^{Al_xPt_y}_{bulk} - xF^{Al}_{bulk} - y(F^{Pt}_{bulk} + F^{S.\ Pt}_f)}{y}, \qquad (7)$$

where $c_s$ is the solubility limit of Pt in Al in at. % as a function of temperature T. $\Delta F^{S.}_{excess}$ is the excess free energy per solute atom, $F^{Al_xPt_y}_{bulk}$ is the Helmholtz free energy of the first ordered phase, and $k_B$ is the Boltzmann constant. Since the solubility limit depends on the first ordered phase, we have chosen both Al4Pt and Al5Pt as they are the 2 possible first ordered phases in the phase diagram as discussed in the previous section.

Figure 4 shows the calculated solubility limit for Pt in Al matrix, with and without the entropy contribution. It can be observed that the calculated solubility limit is close to the experimental value at the temperature slightly higher than the eutectic temperature when considering Al5Pt as the first stable Al-rich phase (see Figure 4) [9]. This also suggests that Al5Pt could be the first ordered phase in the Al-Pt system as indicated in the previous section.

To quantitatively analyse the solute formation energies and the electronic interactions between Pt and Al, we calculate the bonding electron density as

$$\Delta \rho = \rho - \rho_{IAM}, \qquad (8)$$

where $\rho$ is the electron density determined by a self-consistent calculation after full relaxation, and $\rho_{IAM}$ is the electron density based on the independent atom model. Figure 5 presents the bonding electron density in the (110) plane since the bonding electrons are located in the tetrahedral holes in Al [34]. We also compare electron densities along 3 directions with largest magnitude of bonding electron densities (see Figure 5 (b)). It is evident that Pt forms the strongest and Ag forms the weakest bond to Al among all 4 alloying elements, in the same order as their solute formation energies: $E_f^{s.\ Pt} < E_f^{s.\ Au} < E_f^{s.\ Cu} < E_f^{s.\ Ag}$.

**Table 2.** Solute formation energy ($E_f^{s.}$) and solubility at eutectic temperatures ($c_s^{eutctic}$) of Pt, Cu, Au and Ag in the Al matrix. The eutectic temperatures are 930K, 830K, 923K, and 840K for Pt, Cu, Au, and Ag respectively. For Cu, Au, Ag solutes, our results agree well with previous work. These elements have much higher solute formation energy than Pt in Al matrix.

| Elements | $E_f^{s.}(eV)$ | $c_s^{eutctic}$ (at. %) |
|---|---|---|
| Pt | - 1.90 | 0.44 [b] |
| Cu | - 0.15<br>- 0.08 [a] | 2.48 [c] |
| Au | - 0.57<br>- 0.58 [a] | 0.06 [d] |
| Ag | 0.07<br>0.02 [a] | 23.5 [e] |

[a] DFT calculations reported in reference [31], [b, c, d, e] obtained from the experimental phase diagrams from [9], [16], [17], [35] respectively.

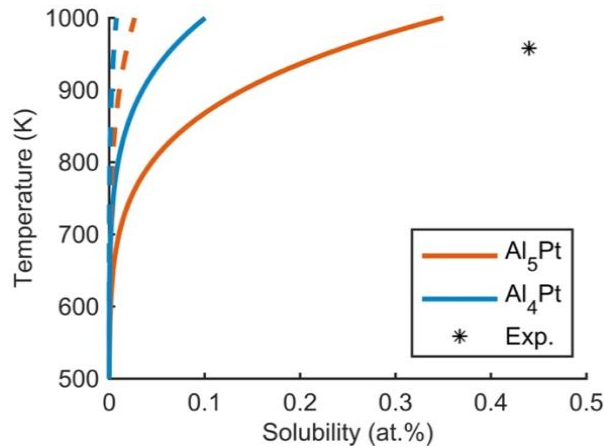

**Figure 4.** Calculated solubility limit of Pt in Al when considering Al$_5$Pt or Al$_4$Pt as the first ordered phase. Dashed lines are solubility limit calculated without the entropy contribution. The star marks the experimentally measured solubility (0.44 at%) of Pt at the eutectic point.

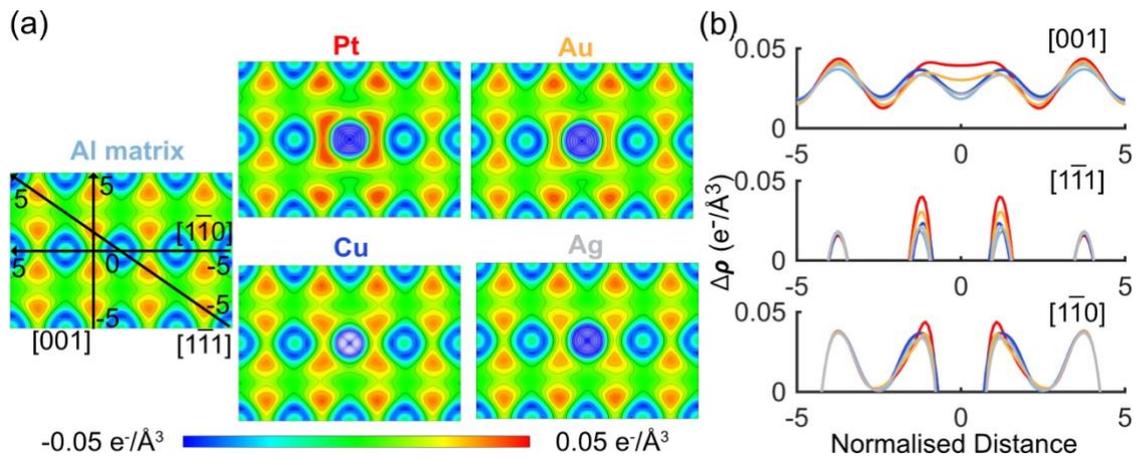

**Figure 5.** Bonding electron density in Al matrix with substitutional solute atoms. In each case, the bonding electron density is plotted on (110) plane passing through the solute atom. The bonding electron density in pure Al matrix is also shown for reference. (b) The distribution of the bonding electron density along 3 directions with largest magnitude of bonding electron densities. Colours for curves are associated with the colours of the titles in (a). The highest bonding electron density is observed for Pt solute.

## 3.3 Pt Clusters and GP Zone like Structures in Al Matrix

Having investigated the thermodynamics of Pt solute in Al matrix, the next step is to determine whether clusters or GP zones will form in the early stages of precipitation, as these will affect the subsequent formation of precipitate phases [2]. To this aim, we model various cluster configurations by considering 3 substitutional Pt/Cu atoms in a large $4 \times 4 \times 4$ Al supercell. In the Al-Cu system, a GP zone is a single Cu layer in Al matrix and $\theta''$ (GP-II zone) is formed by two Cu layers separated by 3 Al layers in Al matrix. Given the similarity between the Al$_2$X phases in Al-Pt and Al-Cu systems, GP zone like structures in the Al-Pt system are modelled as one or two infinitely large and monoatomically thin Pt layers on the {002} planes and separated by several $\{002\}_{Al}$ layers. The supercell is sufficiently large to represent both GP zone-like structures and $\theta''$ like structures. For all configurations, we calculate the excess energy per solute atom relative to Pt solute as

$$E_{excess}^{S.} = \frac{E_{bulk}^{Al_xPt_y} - xE_{bulk}^{Al} - y(E_{bulk}^{Pt} + E_{f.}^{S.\ Pt})}{y}. \tag{9}$$

We also include the data for Cu solute atoms for a comparison with Pt solute atoms. The calculated excess energies for all configurations are shown in Figure 6. We find that the energetically most favourable cluster configurations consist of three nearest neighbour Pt atoms located on (001) or (111) planes (see Figure 6 (a)). These planes are the most common precipitate habit planes in Al alloys [2]. Whereas three coplanar Pt atoms can exhibit a negative excess energy, a single Pt layer in Al matrix shows positive excess energies both on (001) and (111) planes (0.07 eV and 0.36 eV, respectively). The excess energies are positive for all configurations of GP zone like structure for Pt (see Figure 6 (b)). Two Pt layers on (001) planes separated by 7 layers of Al is more favourable than other configurations. In contrast, two Cu layers on (001) separated by 3 layers of Al is more favourable than other configurations (i.e., identical to θ″ phase). When compared with Pt, Cu has negative excess energies in all configurations of clusters (Figure 6 (a)), GP zones, and θ″ [36].

Based on these results, we conclude that coherent precipitates such as GP zones and θ″ are unlikely form in the Al-Pt binary alloy system. Large solute clusters of Pt are also unlikely to form since only a limited number of stable configurations exist. It should be noted that although entropy is not considered here, the overall conclusion about clusters, GP zones and θ″ in Al-Pt systems will not change as the entropy have a positive contribution to the excess energy for all configurations.

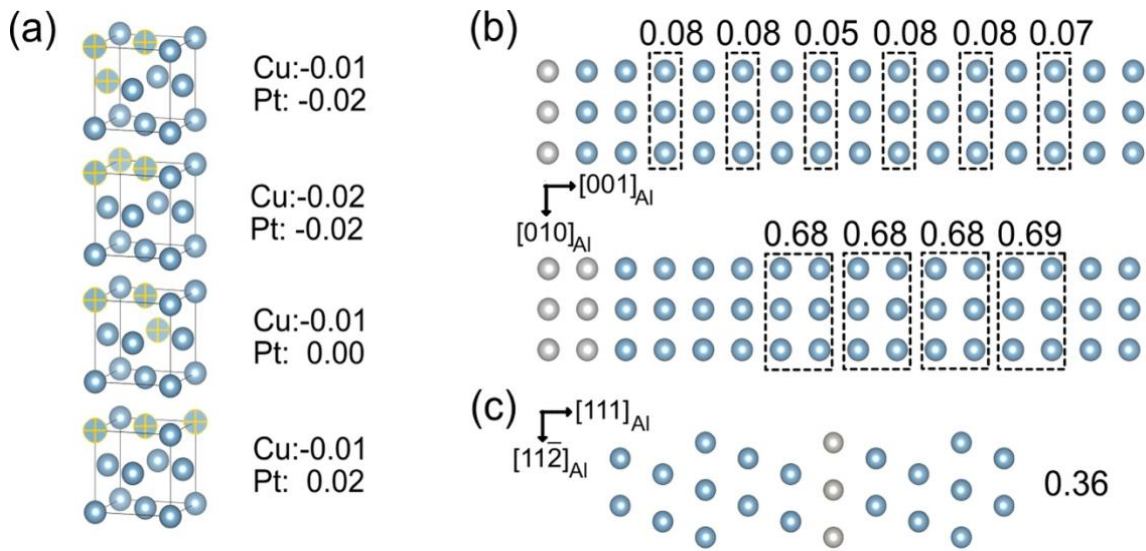

**Figure 6.** Calculated $E_{excess}^{s.}$ (eV) for (a) Cu and Pt clusters, Pt layers on (b) $(001)_{Al}$ planes and (c) on a $(111)_{Al}$ plane. Al and Pt atoms are shown as blue and grey spheres, respectively. In (a), the position of Cu or Pt atoms are marked in yellow. In (b), the position of the second one or two infinitely large Pt layers is marked as dashed rectangles. The excess energy remains positive regardless of the distances between Pt layers, indicating GP zones are unlikely to form in Al-Pt alloys.

### 3.4 Precipitate Shape and Crystallographic Relationship with Matrix

Precipitate shape and their crystallographic orientation in the matrix have a strong influence on mechanical properties of the binary alloy—high-aspect-ratio precipitates on low-index planes such as {001} or {111} tend to be most efficient at blocking stress-induced dislocations and thus imparting high strength to the alloy [2]. Such precipitates are usually coherent with the matrix in one or two dimensions [2], which means some of their interfaces or edges match well with the matrix. It is therefore critical to determine the possible crystallographic relationship(s) between precipitates and matrix, which will determine the precipitate shape and interfaces. To do so, we adopt Kelly et al.'s method [37, 38] whereby low-index crystallographic planes in Al and precipitate structures are compared. The search was limited to the three most densely packed planes for each crystal structure, as they are more likely to result in a good match and hence low-energy interfaces. More specifically, we looked for cases where both interplanar and interatomic spacings have less than 10% misfit [37], and consequently, a greater likelihood for the formation of a coherent interface. Only $Al_2Pt$, $Al_3Pt$ and $Al_5Pt$ were found to satisfy these matching conditions. Figure 7 shows the matched planes and corresponding interfaces between $Al_2Pt$ and Al matrix. The best match for $Al_2Pt$ in Al is for the following crystallographic relationship: $(002)_{Al_2Pt}||(002)_{Al}$ and $(220)_{Al_2Pt}||(200)_{Al}$, leading to a coherent interface, $(002)_{Al_2Pt}||(002)_{Al}$, and 2 sets of semicoherent interfaces, perpendicular to the coherent interface. The coherent interface is likely to be much larger than the semicoherent interfaces. This is analogous to the case of $\theta'$ ($Al_2Cu$) and $\eta$ ($Al_2Au$) precipitates in Al [3, 39].

Based on the orientation relationship determined above, we constructed supercells representing an $Al_2Pt$ precipitate embedded in Al matrix to examine the stability of different interfaces using DFT calculations. When lattice parameters $a$ and $b$ of Al$_2$Pt embedded in Al matrix are fixed to $a_{Al} = 4.04$ Å to model the coherent interface, the lattice parameter $c_{Al_2Pt/Al}$ is found to increase from 5.94 Å in bulk to 6.27 Å in the Al matrix. As shown in Table 3, the formation energy per atom and excess energy per solute are both negative, suggesting a supersaturated solid solution of Pt in Al can be expected to decompose into stable Al$_2$Pt precipitates in the above orientation relationships. Because non-bulk like interfaces have been observed within the θ′ (Al$_2$Cu) structure as a result of Cu solute segregation, the segregation of Pt into the interfaces of Al$_2$Pt is considered [40]. Segregation of a Pt atom increases the formation energy by 0.70 eV, indicating the Al$_2$Pt will displays a bulk like interface similar to η (Al$_2$Au) rather than θ′ [40, 41]. This is mainly due to the significantly lower solute formation energy of Pt in Al compared to that of Cu. The entropy is not considered when examining interfaces as the embedded precipitate in our model is small compared to the Al matrix.

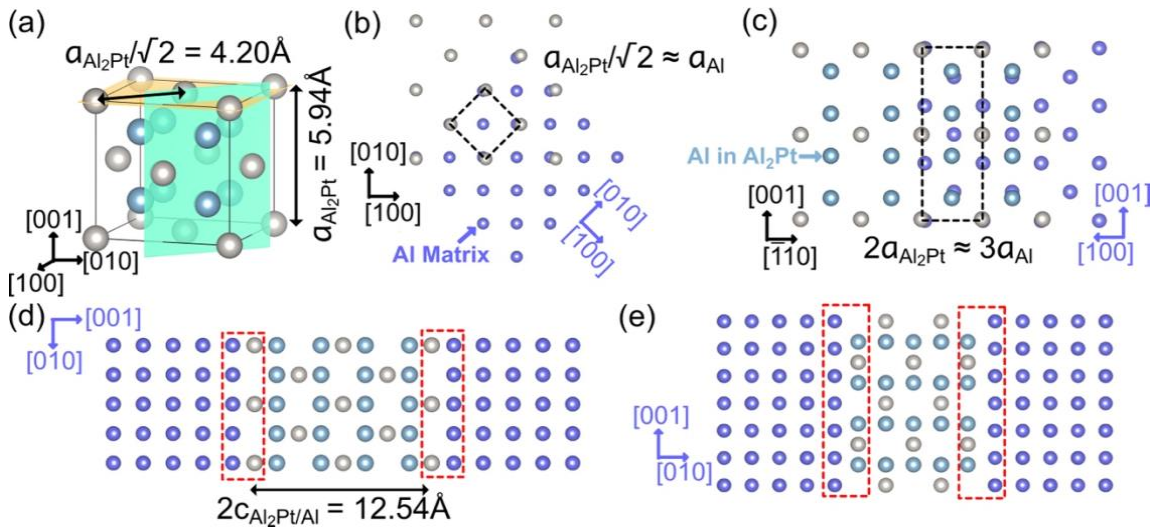

**Figure 7.** DFT relaxed crystal structures of (a) cubic Al$_2$Pt with the marked matched planes, (b) matched planes of $(002)_{Al_2Pt}||(002)_{Al}$ (c) matched planes of $(220)_{Al_2Pt}||(200)_{Al}$. Two stable interfaces are found between Al$_2$Pt precipitates and Al matrix: (d) a coherent interface and (e) a semicoherent interface. Interfaces are marked by red rectangle.

**Table 3.** Formation energy ($E_f^{at.}$) and excess energy ($E_{excess}^{s.}$) for Al$_2$Pt precipitates in the Al matrix as calculated by DFT. The negative energy values indicate that Al$_2$Pt precipitates are thermodynamically stable in the orientation relationship indicated in Figure 7.

| Thickness | $E_f^{at.}$ (eV/atom) | $E_{excess}^{s.}$ (eV/atom) |
|---|---|---|
| 1 $c_{Al_2Pt}$ | -0.20 | -0.05 |
| 2 $c_{Al_2Pt}$ | -0.30 | -0.07 |

The matching of Al$_3$Pt and Al$_5$Pt crystal structures with the Al matrix was analysed with the same method. We found that Al$_3$Pt and Al$_5$Pt precipitates are unlikely to form coherent interfaces because of the positive excess energies (see Supplementary Information 3). In summary, Al$_2$Pt precipitates are likely to have a plate shape because of its 2-dimensional matching to the Al matrix. This is structurally analogous to η (as already mentioned) and to the θ′ phase [3]. Al$_3$Pt and Al$_5$Pt are likely to have equiaxed shapes since no low-energy interface is found, which is also the case for Al$_4$Pt and Al$_{21}$Pt$_8$ [42].

## 3.5 Interfacial Energies for Al$_2$Pt Precipitate in Al Matrix

Interfacial energy is a key parameter controlling the shape and thermodynamics of precipitation. To calculate the interfacial energy, we first calculate the formation energies of Al$_2$Pt precipitate supercells with different size (see Figure 8). Then, the interfacial energies between a Al$_2$Pt precipitate and the Al matrix can be obtained from the slope of the change in formation energy as a function of supercell size [43, 44]. We use the bulk-like interfaces for θ′ (Al$_2$Cu) analogous to that of Al$_2$Pt for an equivalent comparison. It can be seen that Al$_2$Pt has a significantly smaller interfacial energy of 129 mJ/m$^2$ of the coherent interfaces, but a similar interfacial energy of the semicoherent interfaces when compared to θ′ precipitate in Al matrix. Therefore, Al$_2$Pt precipitates may have a higher equilibrium aspect ratio than θ′ precipitate in Al matrix.

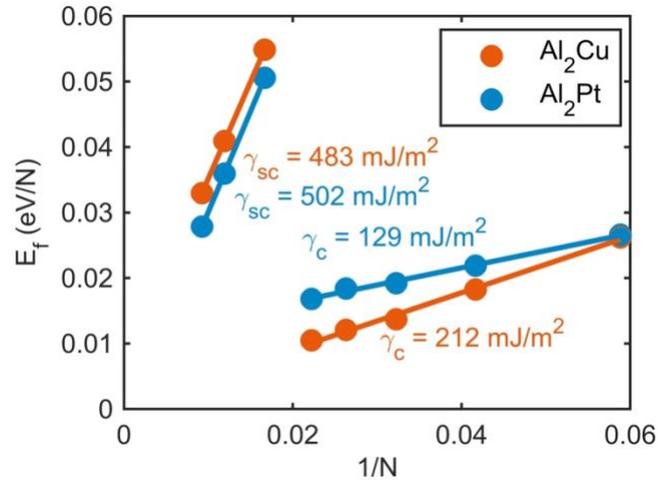

**Figure 8.** Interfacial energy of coherent ($\gamma_c$) and semi-coherent ($\gamma_{sc}$) interfaces of θ′ (Al$_2$Cu) and Al$_2$Pt in Al matrix. N is the total number of atoms of the precipitate supercell and $\gamma$ is calculated from the slope of the fitted line. The coherent interfacial energy of Al$_2$Pt is only a half of that of θ′.

## 3.6 Elastic Properties

The strain energy associated with the misfit of all precipitate phases with Al matrix is the last parameter required as input into our CNT calculations. The elastic properties for all Al-Pt binary phases considered here are calculated based on DFT [45]. Strains within the range of ±1.5% with an increment of 0.5% were applied in all effective directions according to the crystal symmetry of the binary phase. Our results, presented in Table 4, are in good agreement with published values [46], but display appreciable differences with calculations reported in [11], especially for the shear modulus and Young's modulus of $Al_4Pt$ and $Al_2Pt$. The validity of our results is supported by the fact that elastic properties calculated for pure Al are closer to the experimental measurements in our work than in [9].

**Table 4.** Elastic properties including shear modulus ($\mu$), Poisson's ratio ($\nu$), Young's modulus ($E$) and bulk modulus ($K$) of selected phases as calculated by DFT, and compared to reported values. Our results are in good agreement with the reported values.

|  | Al | $Al_5Pt$ | $Al_4Pt$ | $Al_3Pt$ | $Al_{21}Pt_8$ | $Al_2Pt$ |
|---|---|---|---|---|---|---|
| $\mu$ (GPa) | 25 | 59 | 78 | 43 | 62 | 39 |
|  | 24[a] | 76[b] | 34[b] |  | 62[a] | 39[a] |
|  | 35[b] |  |  |  | 64[b] | 64[b] |
|  | 26[c] |  |  |  |  |  |
| $\nu$ | 0.35 | 0.26 | 0.23 | 0.35 | 0.30 | 0.36 |
|  | 0.37[a] | 0.21[b] | 0.36[b] |  | 0.30[a] | 0.36[a] |
|  | 0.30[b] |  |  |  | 0.29[b] | 0.27[b] |
|  | 0.35[c] |  |  |  |  |  |
| $E$ (GPa) | 68 | 148 | 192 | 116 | 162 | 107 |
|  | 66[a] | 182[b] | 93[b] |  | 162[a] | 106[a] |
|  | 91[b] |  |  |  | 165[b] | 182[b] |
|  | 70[c] |  |  |  |  |  |
| $K$ (GPa) | 77 | 103 | 117 | 127 | 133 | 130 |
|  | 83[a] | 104[b] | 112[b] |  | 133[a] | 129[a] |
|  | 74[b] |  |  |  | 133[b] | 130[b] |
|  | 76[c] |  |  |  |  |  |

[a] DFT values reported in [46] and [11] respectively, [c] experimental values reported in [47]

## 3.7 Thermodynamics of Precipitate Nucleation

Having calculated the interfacial energies and strain energies, we can now predict the energy change during nucleation, which measures the likelihood of precipitation of various phases in the Al-Pt binary system. We have assumed a plate-like shape for the $Al_2Pt$ precipitate and a spherical shape for the remaining binary phases with incoherent interfaces. By assuming homogeneous nucleation directly from a supersaturated solid solution, and $\gamma = \gamma_{semi}^{Al_2Pt}$ for all incoherent spherical phases, we calculate the thermodynamics of precipitate nucleation based on the CNT formalism.

Figure 9 shows the change in Gibb's free energy for nucleation of phases in Al-Pt binary system. Among all phases considered here, $Al_2Pt$ has the lowest nucleation energy barrier and critical radius, indicating it will be the first phase to precipitate, in agreement with the experimental observation [19]. Among the three smallest thicknesses of $Al_2Pt$ nuclei that we considered, $1\ c_{Al_2Pt}^{embedded}$ is the most favoured thickness according to our calculations. This is because the large, favourable chemical potential for $Al_2Pt$ to nucleate outweighs the unfavourable contribution of misfit strain energy due to a greater thickness (see Supplementary Information 4). Also worth noting is the much smaller critical radius and barrier to nucleation for $Al_2Pt$ compared to $\theta'$ ($Al_2Cu$) (see Supplementary Information 4). Again, this is mainly due to the significantly larger chemical potential for $Al_2Pt$ to nucleate compare with that of $\theta'$. As displayed in Table 5, the nucleation temperature has a negligible effect on the ranking of the nucleation related parameters. Based on these calculations, the phases can be ranked in terms of ease of nucleation as: $Al_2Pt > Al_{21}Pt_8 > Al_4Pt > Al_5Pt > Al_3Pt$, with $Al_2Pt$ being the easiest phase to nucleate from the solid solution.

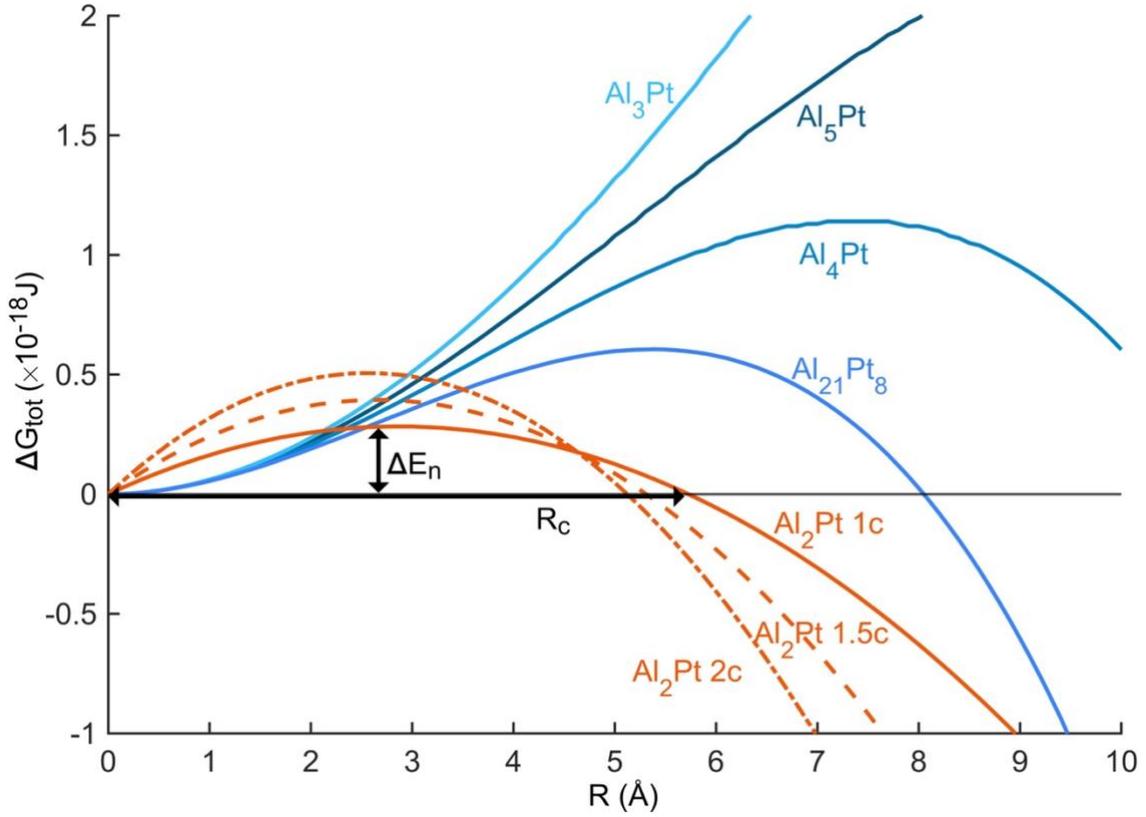

**Figure 9.** Change in Gibbs free energy for the nuclei of various phases in Al-Pt binary system at 500K, as calculated by CNT. Al$_2$Pt with different thickness are plotted in solid line (1c), dashed line (1.5c), and dash-dotted line (2c) respectively. The nucleation energy barrier ($\Delta E_n$) and critical radius ($R_c$) of Al$_2$Pt with the smallest thickness (1c) are marked.

**Table 5.** Nucleation energy barrier ($\Delta E^n$), critical radius ($R_c$), and critical number of Pt atoms ($n_{req}^{Pt}$) for nucleation at 300K and 500K.

| Nucleus | $\Delta E^n$ ($\times 10^{-18} J$) | | $R_c$ (Å) | | $n_{crit}^{Pt}$ | |
|---|---|---|---|---|---|---|
| | 300K | 500K | 300K | 500K | 300K | 500K |
| Al$_5$Pt | 2.02 | 2.32 | 10 | 11 | 4 | 5 |
| Al$_4$Pt | 1.01 | 1.14 | 7 | 7 | 3 | 3 |
| Al$_3$Pt | 10.6 | 8.48 | 23 | 20 | 35 | 28 |
| Al$_{21}$Pt$_8$ | 0.503 | 0.566 | 5 | 5 | 2 | 2 |
| Al$_2$Pt – 2c | 0.479 | 0.506 | 2 | 3 | 4 | 4 |
| Al$_2$Pt – 1.5c | 0.327 | 0.394 | 3 | 3 | 3 | 3 |
| Al$_2$Pt – 1c | 0.266 | 0.283 | 3 | 3 | 2 | 3 |

## 4    Discussion

### 4.1    Phases in the binary Al-Pt System

Intermetallic phases with higher Al content than the precipitate phase (i.e., $Al_2Pt$) are also of interest here because they may form in Al-rich Al-Pt alloys. As reflected by studies published to date [7-9, 12, 13, 18, 19, 48], it hitherto remains unclear which Al-Pt phases are stable or metastable. The full crystal structure of these phases is also not known. For example, the stable phase with the highest Al concentration is believed to be $Al_{21}Pt_5$ [9, 12, 13], even though its full crystal structure is yet to be determined. In disagreement with this report, a recent experimental work claimed that the equilibrium phase with the highest Al content might be $Al_4Pt$ [7, 8]. Our work supports this view—$Al_4Pt$ is found to have a more negative formation energy per atom (-0.54 eV) compared to $Al_{21}Pt_5$ (-0.36 eV), and sits close to the convex hull (see Figure 2). Therefore, our work strongly suggest $Al_4Pt$ is the most likely equilibrium phase with highest Al content.

When exploring the Al-rich part of the Al-Pt phase diagram, we considered the possible phase $Al_5Pt$ (see Section 3.1), as it may compete with $Al_4Pt$ for precipitation. A metastable phase $\epsilon$ with a composition of $Al_5Pt$ was reported to precipitate first in rapidly quenched Al-Pt alloys. This phase ($\epsilon$) was determined to have an atomic volume of 17.05 $Å^3$ [18]. As shown in Section 3.1, we proposed a crystal structure for $Al_5Pt$ (I4/mcm) and calculated its formation energy, which was found to be negative and to be slightly above the convex hull similar to previous calculation [11]. Our proposed crystal structure results in an atomic volume of 17.09 $Å^3$, which is very close to the experimental value [18]. Therefore, $Al_5Pt$ can be considered a plausible candidate for precipitation.

A metastable phase of composition Al$_3$Pt has also been suggested as a possible precipitate phase at room temperature [19, 20]. More recently, a high-temperature stable phase denoted $\xi$ and of the same composition Al$_3$Pt was reported to be stable [7]. Based on DFT calculations for several possible archetypal structures, we proposed a crystal structure for Al$_3$Pt with space group Pm$\bar{3}$n (see Section 3.1). For Al$_3$Pt in our work, we used the most stable structure (i.e. Pm$\bar{3}$n) because the full crystal structure of $\xi$ is still unknown [7]. However, the calculated XRD powder pattern for Al$_3$Pt (Pm$\bar{3}$n) (see Supplementary Information 1) does not match the XRD pattern of $\xi$, indicating that the phase characterised in Ref. [7] does not have our proposed structure. Metastable Al$_6$Pt was also reported in quenched Al-Pt samples [19]. The only crystallographic information available is that Al$_6$Pt is isostructural with the Ga$_6$Pt structure with lattice parameters $a_{Al_6Pt} = 15.762$ Å, $b_{Al_6Pt} = 12.103$ Å and $c_{Al_6Pt} = 8.318$ Å [19], but their detailed atomic positions remain unknown [48]. In summary, our work confirms the equilibrium phase is Al$_4$Pt rather than Al$_{21}$Pt$_5$. We also suggest possible crystal structures for Al$_5$Pt and Al$_3$Pt.

### 4.2 Precipitation Sequence in Al-Pt

Several studies have examined precipitation in Al-Pt alloys [7, 8, 18-20]; however, all but one [19] deal with precipitation from the melt rather than solid solution. In this section, we focus on solid-state precipitation, which is important for high-strength properties. First, no GP zones have been reported to form in Al-Pt alloys, and our DFT results presented in Section 3.3 also suggests that large clusters and GP zones of Pt are thermodynamically unstable in Al.

We predict the $Al_2Pt$ precipitate phase to adopt an orientation relationship similar to that of θ′ and η phase in Al-Cu and Al-Au binary systems respectively. This is different from what was proposed in [19] based on electron diffraction methods. We found that the orientation relationship in [19] provides poor matching between DFT relaxed $Al_2Pt$ and Al matrix, in contrast to our proposed orientation relationship. As shown in Section 3.7, $Al_2Pt$ appears to be the strongest candidate to precipitate first from the solute solution. This agrees with early studies of solid-state precipitation in Al-Pt alloys by Chattopadhyay et al. [19, 49]. This may seem surprising at first, because $Al_2Pt$ is in competition with as many as four phases with much greater Al content, namely: $Al_6Pt$, $Al_5Pt$, $Al_4Pt$ and $Al_3Pt$. These phases have been observed after rapidly quenching Al-(2-3)at.% Pt alloys from the melt [18-20]. Indeed, the $Al_6Pt$ phase was reported to precipitate from the solute solution, but after $Al_2Pt$ [19]. Given the full crystal structure of $Al_6Pt$ is still not known, it is difficult to accurately calculate the thermodynamic parameters for its nucleation such as nucleation energy barrier and critical radius. Nevertheless, using crude approximations, we find that $Al_6Pt$ is likely not to precipitate prior to $Al_2Pt$ (see Supplementary Information 5). This is in agreement with Ref. [19].

Regarding the $Al_5Pt$ and $Al_3Pt$ phases, they contain lattice planes that match Al matrix planes well. However, the interfaces that were considered based on matching these planes were found to be unstable by DFT. Although no thermodynamically stable interfaces were found in this work, a more systematic search may reveal low energy interfaces. In this case, $Al_5Pt$ and $Al_3Pt$ may form prior to $Al_2Pt$ (see details in Supplementary Information 6). However, this situation would contradict the experimental evidence available to date [18, 20].

## 4.3 Effect of Bonding Electron Density on Solute Formation Energy

Our DFT calculation indicated that Pt has a significant negative solute formation energy in Al. The large bonding electron density between Al and Pt as shown in Figure 5 may explain the unfavourable segregation of Pt into the interfaces of Al$_2$Pt precipitates compared to that of Cu, and the low coherent interfacial energy of Al$_2$Pt compared to θ′ (Al$_2$Cu). Compared to Pt and Cu, the bonding between Al and Ag is weaker than that between Al atoms, which is consistent with the positive solute formation energy of Ag in the Al matrix. The distribution of bonding electron density is different for different solute atoms (Cu, Au, Pt, Ag) — in the case of Pt and Au, bonding is stronger in the <110> directions compared with Cu and Ag. This can be correlated with the solute formation energy which is most negative for Pt and Au.

## 4.4 Workflow to predict solid-state precipitation in an arbitrary binary alloy system

Finally, to predict solid-state precipitation in an arbitrary binary alloy system M-X, where M is the matrix element and X is the solute, we propose the following systematic workflow. The first step is to identify all stable and metastable intermetallic phases known between two chemical elements based on the equilibrium phase diagram and databases of crystal structures [6]. A precipitate phase is likely to be structurally related to known bulk intermetallic phases, but not always, as is the case for θ′ (Al$_2$Cu). The full crystal structure of each phase must be known in order to carry out thermodynamics calculations using DFT methods. Because some phases are only stable in a certain temperature range, we need to consider the temperature effect on their stability by estimating the entropy contribution. A major challenge here is that some intermetallic phases are either unknown or have not had their crystal structure solved. In this case, one approach is to adopt crystal structures of phases in similar systems, such as systems between elements in the same group on the periodic table.

The second step is to determine the solute behaviour of X in M. This allows us to determine properties such as solubility, segregation energy and chemical potential. These properties are required in the subsequent steps.

The third step is investigating the interaction of solute atoms, including the possible formation of clusters and GP zone-like structures. The main challenge here is to consider all possible configurations of solute aggregates. Currently, accurate DFT calculations are limited to hundreds of atoms, and therefore only up to a handful of solute atoms. Nevertheless, these methods allow one to explore whether fully coherent precipitates might exist.

The fourth step is predicting the shape and interfacial structure of precipitate phases, which is a great challenge. This requires the orientation relationship(s) between a given precipitate phase and matrix to be determined. One approach is to match the crystal structure of the precipitate phase with the matrix [37, 38]. Based on the matched planes, we can then propose possible low energy interfaces, calculate their interfacial energies and misfit strain energies, and suggest the possible shape of each precipitate phase.

Finally, the last step is to predict the nucleation barrier and critical size of each precipitate phase, thus allowing the precipitation sequence to be proposed. One approach, as used here, involves CNT calculations using DFT calculated quantities. Our CNT calculations are rather crude with many assumptions especially for the interfacial energies of incoherent precipitates. However, it allows a qualitatively convincing investigation of precipitation in the unexplored Al-Pt alloy system.

## 5   Conclusion

In this work, we propose a workflow to predict the solid-state precipitation behaviour in the binary Al-Pt system, based on a combination of atomistic first principles methods using density-functional theory and classical nucleation theory calculations. The bulk intermetallic phases $Al_4Pt$, $Al_{21}Pt_8$, $Al_2Pt$, $Al_3Pt_2$, AlPt (α & β), $Al_3Pt_5$, $AlPt_2$ (α & β) and $AlPt_3$ (α & β) are confirmed to be thermodynamically stable. Our calculations also support the possible existence of $Al_5Pt$ and $Al_3Pt$. A significantly low formation energy of Pt in Al is found. This results in a high energy barrier for Pt to form clusters, GP zones, and non-bulk interfaces. When considering precipitate phases, our calculations strongly suggest that $Al_2Pt$ will adopt a similar orientation relationship and shape to the θ′ ($Al_2Cu$) phase, and will precipitate first. Similarly to θ′, $Al_2Pt$ precipitates are expected to exhibit plate-like shapes with high aspect ratios. Our results are in partial agreement with previous inconclusive experimental work [19, 49], and motivate further experimental validation through a detailed microscopic characterisation of solid-state precipitation in Al-Pt alloys. Our work constitutes an initial step towards predicting solid-state precipitation in an arbitrary Al alloy system, and ultimately, towards the discovery of high-performance Al alloys.

# 6  Acknowledgement

This research was funded by the Australian Government through an Australian Research Council Discovery Project grant (DP210101451). Computational work was undertaken with the assistance of resources and services from the National Computational Infrastructure (NCI), Pawsey Supercomputing Centre, and from the MonARCH HPC cluster.

Supplementary Information

# Predicting the Early Stages of Solid-State Precipitation in Al-rich Al-Pt Alloys

Shenghan Su [a], Laure Bourgeois [a, b, *], Nikhil V. Medhekar [a, *]

[a] *Department of Materials Science and Engineering, Monash University, VIC, 3800, Australia.*

[b] *Monash Centre for Electron Microscopy, Monash University, VIC, 3800, Australia.*

[*] *Corresponding authors.*

*E-mail: nikhil.medhekar@monash.edu; laure.bourgeois@monash.edu*

# S1. Details of Classical Nucleation Theory (CNT) Calculations

To predict the precipitation sequences, the energy change during nucleation of different precipitate phases is calculated according to the classical nucleation theory (CNT) as

$$\Delta G_{tot} = V\Delta G_{chem} + V\Delta G_{el} + A\gamma, \tag{S1}$$

where $V$ is the volume of the nucleus, $\Delta G_{chem}$ is the chemical energy, $\Delta G_{el}$ is the elastic strain energy, $\gamma$ is the interfacial energy, and $A$ is the area of the interfaces between the precipitate phase and matrix. $\Delta G_{chem}$ for a precipitate phase can be estimated by its excess Helmholtz free energy per atom, $\Delta F_{excess}^{at.}$, at a certain temperature $T$:

$$\Delta G_{chem}(T) = \Delta F_{excess}^{at.}/V_{atom}, \tag{S2}$$

$$\Delta F_{excess}^{at.} = \Delta E_{excess}^{at.} - T S_{excess}^{at.}, \tag{S3}$$

where $V_{atom}$ is the atomic volume, $\Delta E_{excess}^{at.}$ is the excess energy per atom calculated at 0K and $S_{excess}^{at.}$ is the entropy contribution per atom at a certain temperature. Calculations at 300K and 500K are carried out to represent precipitation at the room temperature and the aging temperature. Because all precipitates are embedded in the Al matrix, $V_{atom}$ is calculated by the atomic volume of Al as

$$V_{atom} = (a_{Al})^3/4 = 16.5 \text{ Å}^3. \tag{S4}$$

For plate-like Al₂Pt, the elastic energy can be calculated based on the Christian's approximation [1, 2]:

$$\Delta G_{el}^{plate} = \frac{\mu}{1-\nu}\frac{\pi}{4}\varepsilon^2\frac{t}{2R}, \tag{S5}$$

where $\mu$ is the shear modulus, $\nu$ is the Poisson's ratio, and $\varepsilon$ is the tensile strain normal to habit planes. Since Al and Al₂Pt have very similar shear modulus, $\mu$ is assumed to be the same for matrix and precipitates (25 GPa). The strain along the coherent direction can also be ignored as it has little effect compared to the strain along the semicoherent direction. The tensile strain can then be evaluated as [2]

$$\varepsilon = 2\frac{nc_{ppt.} - ma_{matrix}}{nc_{ppt.} + ma_{matrix}}. \quad (S6)$$

For spherical phases that are incoherent with the matrix, there is no constraint to match the lattices and lattice sites between the precipitates and matrix. Therefore, Nabarro's approximation can be used to calculate the elastic strain energy of a spherical inclusion of the unstrained radius volume as [3]

$$\Delta G_{el}^{sphere} = \frac{6E\varepsilon^2}{1+\frac{4E}{3K}}, \quad (S7)$$

where $E$ is Young's modulus of the matrix, $K$ is the bulk modulus of the precipitates, and $\varepsilon$ is the strain. Assuming at least one unit cell of the precipitate phase needs to nucleate to represent such phase, a minimum $\varepsilon$ can be calculated as

$$\varepsilon = 1 - \text{Min}(|\frac{max(a,b,c)_{ppt.}}{ma_{matrix}}|), \quad (S8)$$

where $max(a,b,c)_{ppt.}$ is the maximum of lattice parameters of the precipitates, and $m$ is an integer from 1. The required number of Pt atoms are also calculated in terms of the critical size of the nucleus

$$n_{crit}^{Pt} = \left\lceil V_{crit} \frac{n_{Pt}}{V_{unit}} \right\rceil, \quad (S9)$$

where $V_{crit}$ is the critical volume of the precipitates calculated based on the critical radius, and $\frac{n_{Pt}}{V_{unit}}$ is the number of Pt atoms per unit of volume in the corresponding intermetallic phase.

**S2. Crystallographic Information and Thermodynamics of Metastable Phases**

Here we investigate phases $Al_5Pt$, $Al_{21}Pt_5$ and $Al_3Pt$, whose crystal structures are not fully known. For these phases, crystal structures are proposed based on intermetallic phases between elements in the same group of Al and Pt.

Table S1 lists the calculated formation energies of different phases. The crystal structures of these phases are adopted from intermetallic phases between elements in the same group of Al and Pt based on the Materials Project database [4]. Amongst two possibilities, the I4/mcm structure is found to be the most stable one for $Al_5Pt$ and is selected for further calculations. For $Al_3Pt$ phase, the $Pm\bar{3}n$ structure is found to have the lowest energy. Because $Al_3Pt$ ($Pm\bar{3}n$) is stabilised by entropy at high temperatures, entropy contributions are considered for all possible crystal structures of $Al_3Pt$, see Figure S1. It can be observed that the $Pm\bar{3}n$ structure remains to be the most stable one. A high-temperature stable phase denoted $\xi$ and of a composition close to $Al_3Pt$ was reported to be stable above 801 °C [5]. Our newly proposed $Al_3Pt$ ($Pm\bar{3}n$) phase is also stable at high temperatures. Therefore, we calculate the XRD powder patterns of all possible structures of $Al_3Pt$. As shown in Figure S2, compared to the experimental measurements [5], $Al_3Pt$ (Pnma) shows a similar noisy pattern to the $\xi$ phase. However, small peaks between 30° and 40° suggests $Al_3Pt$ (Pnma) have a different crystal structure to $\xi$. $Al_3Pt$ ($Pm\bar{3}n$) also shows distinct peaks compared to $\xi$. Therefore, $\xi$ phase may differ from any possible structure of $Al_3Pt$ considered in this work, or it could possibly contain more than one phase. Because the $Pm\bar{3}n$ structure is the most stable one for $Al_3Pt$, and the full crystal structure of $\xi$ has not been solved [5], $Al_3Pt$ ($Pm\bar{3}n$) is used for further calculations in this work.

**Table S1.** Crystal structures and formation energies per atom ($E_f^{at.}$) of possible metastable phases in the Al-Pt system.

| Chemical formula | Space group | Original structure | $E_f^{at.}$ (eV/atom) |
|---|---|---|---|
| Al$_5$Pt | I4/mcm | Ga$_5$Pd | -0.43 |
| | P2$_1$/m | Ga$_5$Pt | -0.40 |
| Al$_{21}$Pt$_5$ | Pm$\bar{3}$m | Ga$_{21}$Ni$_5$ | -0.36 |
| Al$_3$Pt | Pnma | Ga$_3$Pt | -0.48 |
| | Fm$\bar{3}$m | Al$_3$Ni | -0.33 |
| | P6$_3$/mmc | Ga$_3$Pt | -0.45 |
| | I4/mmm | Ga$_3$Pt | -0.44 |
| | Pm$\bar{3}$n | In$_3$Pt | -0.51 |

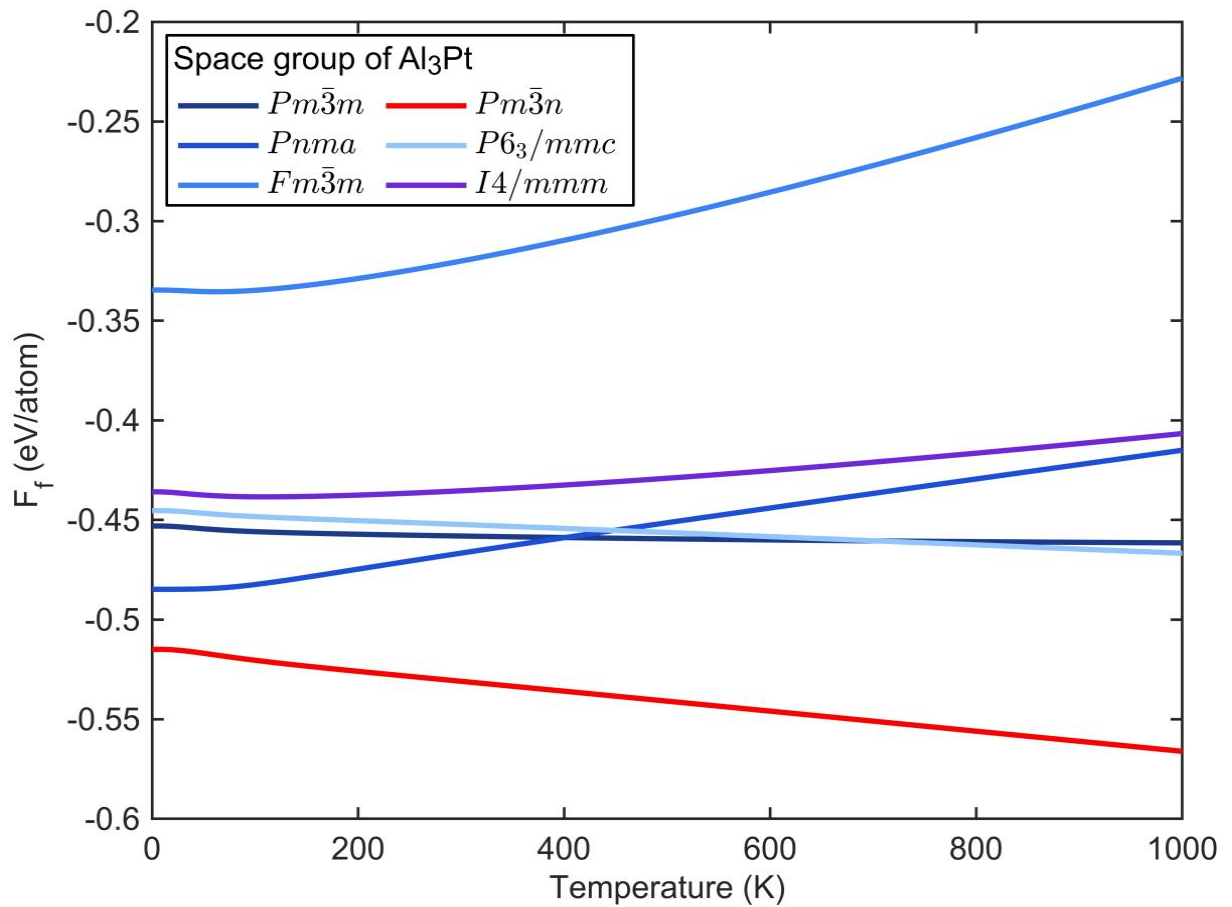

**Figure S1**. Change in free energy of formation per atom from 0K to 1000K for Al$_3$Pt with different crystal structures. The Pm$\bar{3}$n structure remains to be the most stable one for Al$_3$Pt.

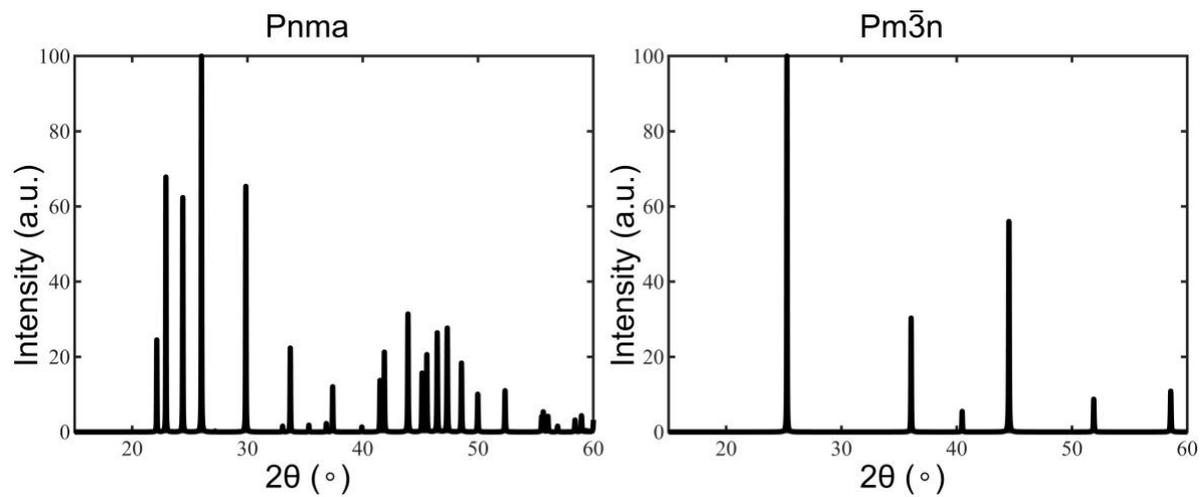

**Figure S2**. Calculated XRD power patterns of Al₃Pt (Pnma) and Al₃Pt (Pm$\bar{3}$n) based on their DFT relaxed crystal structures.

## S3. Interface Matching of Al$_3$Pt and Al$_5$Pt

We adopted Kelly et al.'s method [6, 7] to determine the interfacial structures between a give precipitate phase, Al$_3$Pt and Al$_5$Pt, and Al matrix. Both Al$_3$Pt (Pm$\bar{3}$n) and Al$_5$Pt (I4/mcm) possess certain planes that can match Al matrix (see Figure S3, S4). For Al$_3$Pt, its (200) planes can match the (111) planes of Al. For Al$_5$Pt, its pure Al layers on (002) planes can match the (001) planes of Al. The best matches for these two precipitate phases results in the following crystallographic relationships: $(200)_{Al_3Pt}||(111)_{Al}$ and $(002)_{Al_5Pt}||(001)_{Al}$, leading to interfaces, $(200)_{Al_3Pt}||(111)_{Al}$ and $(002)_{Al_5Pt}||(001)_{Al}$. We then constructed supercells representing Al$_3$Pt and Al$_5$Pt precipitate phases embedded in Al based on the above orientation relationship in order to examine the stability of their interfaces using DFT calculations. The interfaces between Al$_3$Pt/Al$_5$Pt and the Al matrix are not thermodynamically stable due to their positive excess energy $E^{s.}_{excess}$ (see Table S2, S3). Therefore, a spherical shape is assumed for Al$_3$Pt and Al$_5$Pt because there is no stable coherent interface found for these phases.

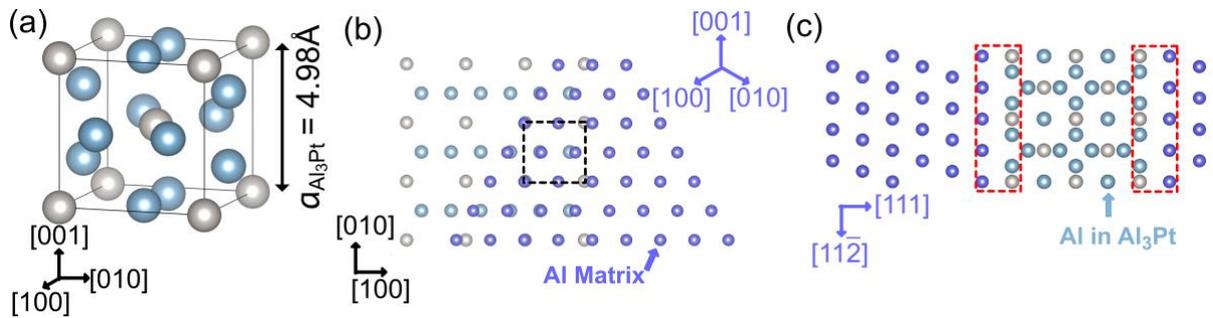

**Figure S3**. Crystal structures of (a) Al$_3$Pt, (b) matched planes of the $(200)_{Al_3Pt}||(111)_{Al}$, and (c) the interfaces between Al$_3$Pt and Al matrix (marked in red rectangular).

**Table S2**. Calculated formation energy per atom ($E_f^{at.}$) and excess energy per solute ($E_{excess}^{s.}$) of Al$_3$Pt precipitates with different thickness. A positive $E_{excess}^{s.}$ indicates thermodynamically unstable interfaces.

| Thickness of Al$_3$Pt | $E_f^{at.}(eV/atom)$ | $E_{excess}^{s.}(eV/atom)$ |
|---|---|---|
| 1 c | -0.06 | 0.01 |
| 2 c | -0.11 | 0.01 |

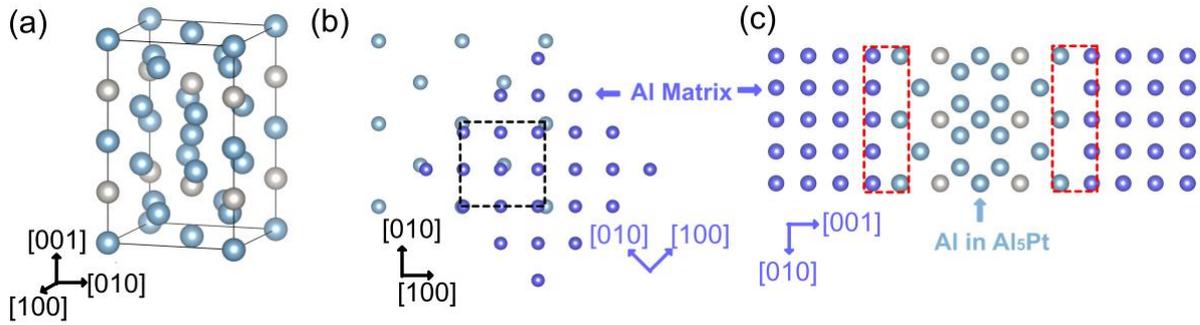

**Figure S4**. Crystal structures of (a) Al$_5$Pt, (b) matched planes of the $(002)_{Al_5Pt}||(001)_{Al}$, and (c) the interfaces between Al$_5$Pt and Al matrix (marked in red rectangular).

**Table S3**. Calculated formation energy per atom ($E_f^{at.}$) and excess energy per solute ($E_{excess}^{s.}$) of Al$_5$Pt precipitates with different thickness. A positive $E_{excess}^{s.}$ indicates thermodynamically unstable interfaces.

| Thickness of Al$_5$Pt | $E_f^{at.}(eV/atom)$ | $E_{excess}^{s.}(eV/atom)$ |
|---|---|---|
| 0.5 c | 0.18 | 0.23 |
| 1 c | 0.29 | 0.38 |

## S4. Comparing the nucleation behaviour of Al$_2$Pt to θ′ (Al$_2$Cu)

Because Al$_2$Pt shows certain similarities to θ′ (Al$_2$Cu), we compared their nucleation behaviour. In Figure S5, dark blue curves and light blue curves are references that show the energy change during nucleation of θ′ and Al$_2$Pt respectively. The phase Al$_2$Pt has much smaller nucleation energy barrier and critical radius compared to θ′. The medium blue curves indicate the contribution of a certain parameter. The large difference in chemical potential is the major effects on nucleation, and the interfacial energy also have noticeable effect. In contrast, the influence of the misfit strain is neglectable.

The thickness of Al$_2$Pt affects precipitation as it leads to different strains: $\varepsilon = -0.25, +0.15, +0.03$ for $t = 1\ c_{Al_2Pt}^{embedded}, 1.5\ c_{Al_2Pt}^{embedded}, 2\ c_{Al_2Pt}^{embedded}$, respectively. A thickness of $2\ c_{\theta'-Al_2Cu}^{embedded}$ is most desired for θ′. However, $1\ c_{Al_2Pt}^{embedded}$ is more favourable than 1.5 or $2\ c_{Al_2Pt}^{embedded}$ for Al$_2$Pt for its lowest nucleation energy barrier and critical number of Pt atoms. This is also because of the large chemical potential for Al$_2$Pt to nucleate outweighs the influence of elastic strain energy.

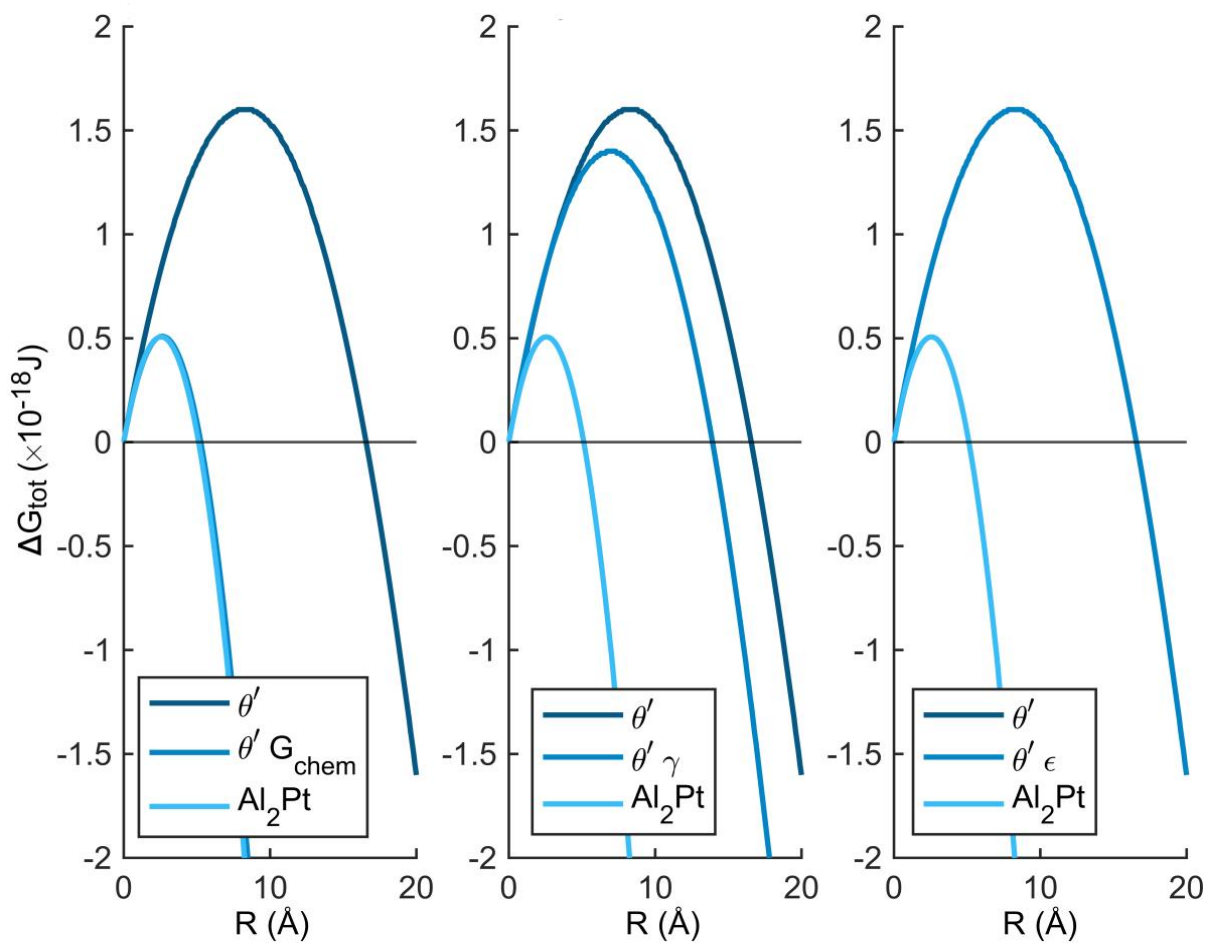

**Figure S5**. Determining influences of CNT parameters on the nucleation energy barrier and critical radius at 500 K. Dark and light blue represent the references of θ′ ($Al_2Cu$) and $Al_2Pt$. In (a), (b), and (c), a single parameter (chemical potential, interfacial energy, strain) of θ′ is changed to the according value of $Al_2Pt$.

## S5. Comparing the nucleation behaviour of Al$_2$Pt to Al$_6$Pt

The Al$_6$Pt phase was reported to precipitate from the solid solution after Al$_2$Pt [8]. Because the full crystal structure of Al$_6$Pt is still unknown, we use crude approximations to estimate its nucleation behaviour. Assuming Al$_6$Pt has the same formation energy per solute atom compared to Al$_2$Pt (i.e., Al$_6$Pt sits on the convex hull), an interfacial energy of 296 mJ/m$^2$ is required to allow Al$_6$Pt has a same nucleation energy barrier to Al$_2$Pt (see Figure S6). This is a small interfacial energy considering the semicoherent interfacial energies of Al$_2$Pt and θ′ (Al$_2$Cu) are around 500 mJ/m$^2$. Because no stable phase has been reported with a stoichiometric of Al$_6$Pt, and an interface with low interfacial energy is undermined, Al$_6$Pt is unlikely to precipitate prior to Al$_2$Pt.

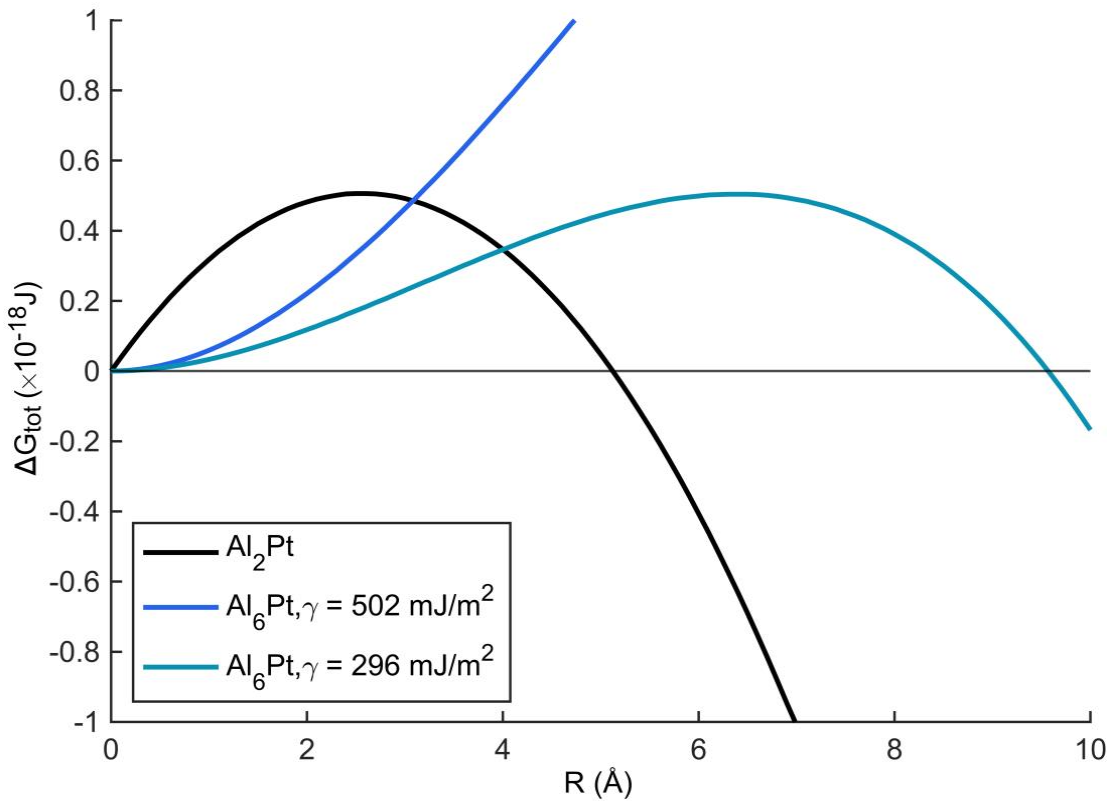

**Figure S6.** Comparing energy change between Al$_6$Pt and Al$_2$Pt during nucleation. The blue curves are spherical Al$_6$Pt with different interfacial energies.

## S6. Comparing the nucleation behaviour of Al$_2$Pt to Al$_5$Pt and Al$_3$Pt

For phases Al$_5$Pt and Al$_3$Pt that contain lattice planes that match Al matrix planes well, possible, yet unrevealed, low energy interfaces may facilitate their precipitation. Following the similar process as in Sec. S5, interfacial energies of 302 mJ/m$^3$ and 196 mJ/m$^3$ are required for Al$_5$Pt and Al$_3$Pt respectively to obtain a same nucleation energy barrier compared to Al$_2$Pt (see Figure S7). These interfacial energies are reasonable when considering their good interface matching. A noticeable obstacle is the larger critical radius of Al$_5$Pt and Al$_3$Pt than that of Al$_2$Pt, which may impede their precipitation from the Pt solid solution.

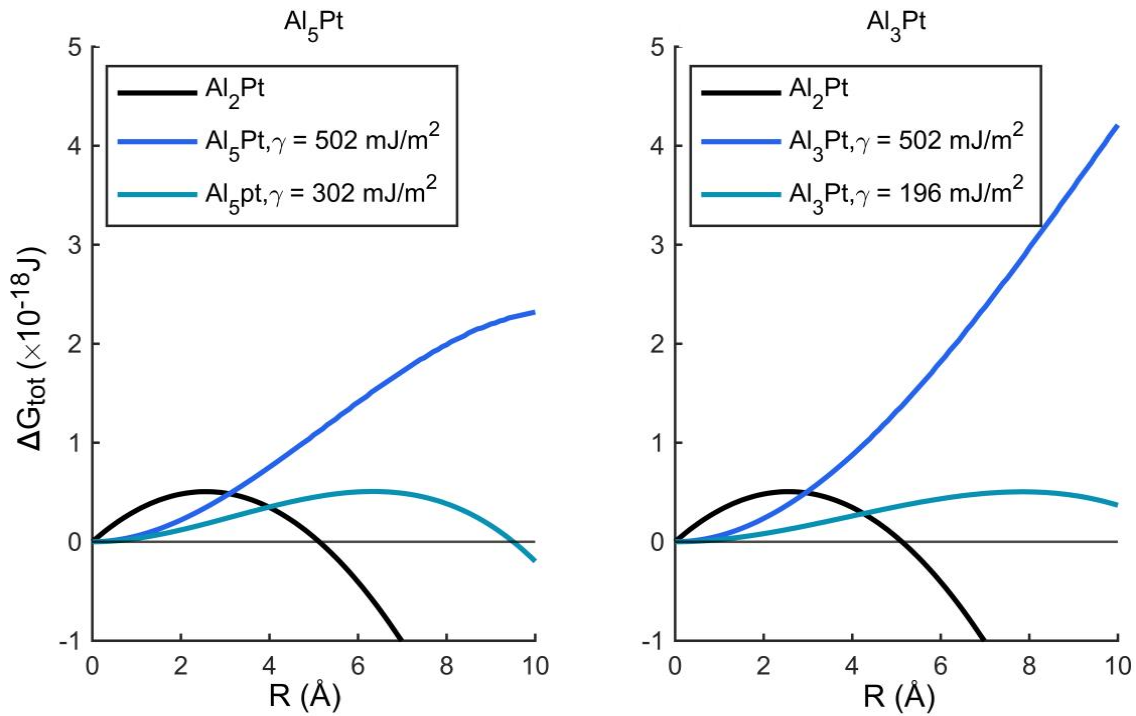

**Figure S7.** Comparing the energy change of a spherical nucleus of Al$_3$Pt and Al$_5$Pt under 500K with different interfacial energies to Al$_2$Pt.